# Closed-Form Rate Outage Probability for OFDMA Multi-Hop Broadband Wireless Networks under Nakagami-m Channels

Mohammad Hayajneh and Najah AbuAli
College of Information Technology
UAE University
Al Ain, United Arab Emirates
e-mail: {mhayajneh, najah}@uae.ac.ae

*Abstract*—Rate outage probability is an important performance metric to measure the level of quality of service (QoS) in the Generation (4G) broadband access networks. Thus, in this paper, we calculate a closed form expression of the rate outage probability for a given user in a down-link multi-hop OFDMA-based system encountered as a result of links' channel variations. The channel random behavior on different subcarriers allocated to a given user is assumed to follow independent non-identical Nakagami-m distributions. Besides the rate outage probability formulas for single hop and multi-hop networks, we also derive a novel closed form formulas for the moment generating function, probability distribution function (pdf), and the cumulative distribution function (cdf) of a product of independent non-identical Gamma distributed random variables (RVs). These RVs are functions of the attainable signal-to-noise power ratio (SNR) on the allocated group of subcarriers. For single-hop scenario, inspired by the rate outage probability closed formula, we formulate an optimization problem in which we allocate subcarriers to users such that the total transmission rate is maximized while catering for fairness for all users. In the proposed formulation, fairness is considered by guaranteeing a minimum rate outage probability for each admitted user. *(Abstract)*

## I. INTRODUCTION

The Worldwide Interoperability for Microwave Access (WiMAX) and Long-Term Evolution (LTE)– also known as Evolved Universal Terrestrial Radio Access (E-UTRA)– are two promising candidates for providing ubiquitous IP 4G wireless broadband access. The LTE and WiMAX are designed as all IP packet switched networks in order to be capable of providing higher link data bit rate [1], [2]. Thus, all services involving real and non- real time data, are implemented on top of the IP as a network layer protocol. Real time data services transmission such as voice and video require maintaining the wireless link data rate within the requirement of the rate demands of these applications. Thus, the outage of the link data rate more than a predefined threshold is not tolerable. The link data rate is said to be in an outage if the rate requirements of an application cannot be satisfied. Hence, we define the rate outage probability as the probability of having the attainable link data rate below the required application data rate. LTE and WiMAX both implement OFDMA as the technique to share the access of the media among users.

OFDMA is a multiple-access scheme that inherits the ability of Orthogonal Frequency Division Multiplexing (OFDM) to combat the inter-symbol interference and frequency selective fading channels. In order to reduce the effect of selective fading channels in wireless networks, OFDMA divides the total frequency bandwidth into a number of narrow-band subcarriers in such a way that each subcarrier (SC) experiences a flat fading channel. This is possible by selecting the SC width to be close to the coherence bandwidth of the channel. A small end-portion of the symbol time, which has duration equal to the maximum delay spread channel, is called a cyclic prefix (CP). The CP is copied at the beginning of the symbol time duration to eliminate the effect of inter-symbol interference (ISI). Thus, the inter-symbol interference is dealt with while orthogonality among the subcarriers is maintained.

Accordingly, OFDMA is a preeminent access technology for broadband wireless data networks compared with traditional access technologies such as TDMA and CDMA. The main advantages of OFDMA over TDMA/CDMA stem from the scalability of OFDMA, the uplink orthogonality and the ability to take advantage of the frequency selectivity of the channel. Other advantages of OFDMA include MIMO-friendliness and ability to support prominent quality of service (QoS).

The rate outage probability defined as the probability of having the attainable link data rate below the required application data rate. Our objective in this work is to derive a closed form expression of the rate outage probability for Orthogonal Frequency Division Multiple Access (OFDMA)-based system under the Nakagami-m channel model. Our interest in calculating the rate outage probability in an OFDMA system is motivated by the fact that OFDMA is the common physical layer technology for the 4th generation (4G) and beyond broadband wireless networks; mainly LTE-Advanced and IEEE802.16m networks. Calculating a closed-form rate outage probability expression facilitates designing sound radio resource management schemes for the efficient operation of these networks in single and multi-hop contexts.

Recently, several attempts are made to compute the outage probability in OFDMA systems. These attempts can be categorized into two major categories: the simulation modeling and the analytical derivation. An example on simulation modeling is the work presented in [3]. Simulation requires collecting sufficient statistics of the network, which is time consuming, then applying a prediction model, which risks various prediction error. Hence, we will focus on the analytical derivation category. The works in [4, 5, 6, 7, 8, 9,and 10] are examples of the analytical derivation attempts for single and multi-hop networks. The authors of [4] and [5] compute uplink and downlink outage probabilities in [4] and [5] respectively under Rayleigh fading channel model. The proposal presented in [6] computes downlink outage probability of all users in a cell for a particular resource allocation (subcarrier). In [7], the author computes the downlink outage probability for a system with a limited channel feedback, and the proposal presented in [8] attempt to describe downlink outage probability related to bandwidth user demand. The work in [10], and [9] addressed the calculation of outage probability in multi-hop networks. The authors of [10] provided three lower bounds of end-to-end outage capacity over Rayleigh fading channels, while [9] provided an approximation on outage probability through the numerical inversion of a Laplace transform over Rayleigh fading channels as well .

The aforementioned analytical proposals, although compute outage probability for different objectives, they do share common procedures. They compute the outage probability for a particular resource allocation (OFDMA subcarrier) and/or employed a simplified channel model such as the Rayleigh channel model. Moreover, all proposals are semi-analytical where outage probability computation is approximated through calculating upper and/or lower bounds or through providing numerical solutions. Nevertheless, none of the above proposals provides a closed form expression of the outage probability. In this paper, we address the void of the current literature by deriving a closed form formula for the rate outage probability in single (i.e., PMP) and multi-hop networks. We compute the outage probability based on a generalized channel model, namely the Nakagami-m model of which Rayleigh is merely a special case. Finally, we express the outage probability formula as a direct function of the user required data rate and two other parameters; the Nakagami-m fading coefficient and the spreading factor.

Our derivation has several applications in 4G and beyond broadband networks. For example, outage probability can be considered as a QoS metric to meet a specific connection data rate requirement, or it can be utilized as a performance measure to evaluate the level of meeting the total demands of users in a cellular network.

The remainder of this paper is organized as follows. In Section 2 we present the problem description and system model, while the computation of the outage probability and the relevant mathematical formulation is offered in Sections 3 and 4. In section 5, we introduce an optimization formulation for single hop cellular system scenario. In Section 6 we conclude the paper.

## II.  SYSTEM MODEL

We consider an OFDMA system model in two network scenarios: namely, single hop cellular radio network, see Fig. 1 and multi-hop cellular radio network, see Fig. 2. Each cell, the network consists of a Base Station (BS) with omnidirectional antenna located in the center of the cell serving multiple Subscriber (SS) stations uniformly distributed over the cell.

The SSs share the total channel bandwidth, which is divided into $N$ orthogonal subcarriers. SSs and BS transmission is on a time frame basis, where each frame consists of multiple OFDMA symbols. The channel condition of the SS is assumed not changing during each time frame. However, it may change from one frame to another. The received low-pass signal at SS $k$ on the $n$th SC at OFDMA symbol $t$ is denoted by $R_{k,n}(t)$ and given as:

$$R_{k,n}(t) = h_{k,n}(t)e^{-jw_{k,n}(t)}S_{k,n}(t) + z_{k,n}(t) \qquad (1)$$

where $S_{k,n}(t)$ is the equivalent low-pass transmitted signal to the $k$ th SS on the $n$ th SC. $h_{k,n}(t)e^{-jw_{k,n}(t)}$ is the time-variant transfer function of the equivalent low-pass frequency non-selective channel of the $k$ th SS experienced on the $n$ th SC. The transfer function $h_{k,n}(t)e^{-jw_{k,n}(t)}$ is a complex-valued Gaussian random process in which $h_{k,n}(t)$ is a Nakagami-m distributed random variable and $w_{k,n}(t)$ is uniformly distributed random variable in $(-f, f)$. $h_{k,n}(t)$ represents the channel envelope (modulus), while $w_{k,n}(t)$ captures the phase of the equivalent low-pass channel. We Adopt the Nakagami-m distribution because it fits the realistic statistical variations of the channel envelope in both line-of-sight (LOS) and non line-of-sight (NLOS) communications. Both, the envelope and the phase of the channel are assumed to be jointly independent. Furthermore, we assume that $h_{k,n}(t)$ and $h_{k,m}(t)$ are independent random variables $\forall m \neq n$, which implies that the channel coherence bandwidth is approximately equal to the bandwidth of a SC. $z_{k,n}(t)$ is the complex-valued white Gaussian noise process with power spectral density denoted by $N_o$.

We denote to the SS's $k$, $(k \lor \{1,...,K\}$ on subcarrier $n$, $n \lor \{1,....,N\})$ data rate by $r_{k,n}(t)$ bits per OFDMA symbol. $r_{k,n}(t)$ depends on the allocated power $p_{k,n}$, SS $k$ on subcarrier $n$, and the channel gain $h_{k,n}$. The channel gain $h_{k,n}$ is modeled as a Nakagami-m random variable with (a probability density function) pdf defined as:

$$f_{h_{k,n}}(\check{S}) = \frac{2m_{k,n}^{m_{k,n}}}{\Gamma(m_{k,n})\Omega_{k,n}^{m_n}} \check{S}^{2m_{k,n}-1} e^{(-\frac{m_{k,n}}{\Omega_{k,n}})\check{S}^2}, \forall n \in \mathcal{N}, (2)$$

where $\Omega_{k,n} = E\{h_{k,n}^2\}$ is a coefficient that reflects the spread of the Nakagami-m distribution and $\mathcal{N} = \{1, 2, \cdots, N\}$ is the indexing set of the data subcarriers. $m_{k,n}$ is the fading coefficient that measures the severity fading of the channel. $m_{k,n}$ is defined as:

$$m_{k,n} = \frac{\Omega_{k,n}^2}{E\{(h_{k,n}^2 - \Omega_{k,n})^2\}}. \quad (3)$$

Note that for $m_{k,n} = \infty$, this corresponds to a non-fading channel. For $m_{k,n} = 1$, the Nakagami-m distribution reduces to the Rayleigh distribution, which is used to model small scale channel variation in NLOS systems, and to one-sided Gaussian distribution for $m_{k,n} = \frac{1}{2}$. Further examples on the generality of the Nakagami-m model is when $m_{k,n} < 1$, in this case Nakagami-m distribution closely approximates the Hoyt distribution, and for $m_{k,n} > 1$, it approximates the Rician channel model. Thus, the Nakagami-m channel model is capable of capturing a realistic wireless channel through the two variables; $m_{k,n}$ and $\Omega_{k,n}$. Hence, Nakagami-m channel is an appropriate model to meet this work objective, which is to facilitate practical resource management through trackable closed-form expression of the rate outage probability.

## III. DERIVATION OF SINGLE-HOP RATE OUTAGE PROBABILITY

In this paper, the measure of the level of QoS satisfaction is given in terms of the attainable data rate, $r_k(t)$, against the SS $k$ minimum required data rate, $r_{k,min}$. Explicitly, a SS is in outage, if the required data rate is not met due to low attainable data rate. The outage of meeting the required data rate can be considered as a QoS metric or as a performance measure. For the case of the outage probability as a QoS metric, a SS can demand a predefined threshold for the probability of the outage of service. The outage probability threshold will be used to control the amount of network resources allocated for the user. Alternatively, the outage probability can be calculated given the allocated resources. In this case the outage probability is considered as a performance measure. For both cases, our aim is to calculate the rate outage probability for a specific SS in consequence of its variable channel quality and adaptive transmission.

The rate outage probability $P_k^{out}$ of the kth SS can be expressed as:

$$P_k^{out} = Pr(r_k(t) \leq r_{k,min}) \quad (4)$$

Now, let us define the achievable transmission data rate by subcarrier $n$ of SS $k$ as [11]:

$$r_{k,n}(t) = \frac{B}{N} \log_2(1 + \mathsf{x}_{k,n}(t)), \quad (5)$$

where $B$ is the link bandwidth and $N$ is the total number of data subcarriers. Total link bandwidth is denoted as $B$ and hence, $\frac{B}{N}$ represents the physical bandwidth of each subcarrier. $\mathsf{x}_{k,n}(t)$ is the Signal to Noise power Ratio (SNR) of SS $k$ at subcarrier $n$ at OFDMA symbol $t$ and is expressed as:

$$\mathsf{x}_{k,n}(t) = \frac{p_{k,n} H_{k,n}(t)}{N_o B / N}, \quad (6)$$

where $p_{k,n}$ is the transmit power of the SS $k$ at the subcarrier $n$, $H_{k,n}(t) \triangleq h_{k,n}^2(t)$ with $h_{k,n}(t)$ the channel gain between the BS and SS $k$ at subcarrier $n$ at OFDMA symbol $t$. $N_o$ is the Additive White Gaussian Noise (AWGN) power spectral density at the receiver. As per our earlier assumption that the channel gains on all subcarriers change form one time frame to another time frame, we drop the time index $t$ in our next calculations. Note that in (6), the interference coming from other users using same subcarriers as SS $k$ in the same OFDMA symbol has been assumed negligible. The impact of the interference in OFDMA signalling can be taken care of in the MAC protocol.

The overall achievable transmission data rate, $r_k$ in its OFDMA symbol, of SS $k$ on all its allocated subcarriers in the subset $\mathcal{C}_k : n = 1, \ldots M_k; M_k \leq N$ is:

$$r_k = \sum_{n \in \mathcal{C}_k} r_{k,n}, \quad (7)$$

$$= \frac{B}{N} \sum_{n \in \mathcal{C}_k} \log_2(1 + \mathsf{x}_{k,n}), \quad (8)$$

$$= \frac{B}{N} \log_2\left(\prod_{n \in \mathcal{C}_k}(1 + \mathsf{x}_{k,n})\right), \quad (9)$$

where $M_k = |\mathcal{C}_k|$, i.e. the cardinality of the subset $\mathcal{C}_k$. Using (9) and (4) the rate outage probability $P_k^{out}$ can be expressed as:

$$P_k^{out} = Pr\left(\prod_{n \in \mathcal{C}_k}(1 + \mathsf{x}_{k,n}) | \mathcal{C}_k \leq 2^{r_{k,min}/B_{sc}}\right), \quad (10)$$

where $B_{sc} \triangleq \frac{B}{N}$ is the bandwidth of each subcarrier. Define

$$X_{k,n} \triangleq 1 + \mathsf{x}_{k,n}, \quad (11)$$

Given that $\mathsf{x}_{k,n}$ is based on a Nakagami-m channel model, it follows that the random variable $X_{k,n}$ has the following Gamma distribution:

$$f_{X_{k,n}}(x) = \frac{(x-1)^{m_{k,n}-1}}{\mathsf{s}_{k,n}^{m_{k,n}} \Gamma(m_{k,n})} \exp(-\frac{x-1}{\mathsf{s}_{k,n}}), \quad (12)$$

where

$$\mathsf{s}_{k,n} = \frac{1}{m_{k,n}} E\{X_{k,n}\} = \frac{1}{m_{k,n}}(1 + \frac{p_{k,n}\Omega_{k,n}}{N_o B_{sc}}).$$

Moreover, define the following new random variable, $Y_k$ as:

$$Y_k \triangleq \prod_{n \in \mathcal{C}_k} X_{k,n}, \quad (13)$$

Thus, the achievable rate of the $k$ th SS $r_k(t)$ is given by:

$$r_k = \frac{B}{N} \log_2(Y_k) \quad (14)$$

and equation (10) is reduced into

$$P_k^{out} = Pr\left(Y_k \mid \mathcal{C}_k \leq 2^{r_{k,min}/B_{sc}}\right), \quad (15)$$

$Y_k$ is a product of independent not necessarily identically (i.n.i.d) Gamma distributed random variables, $X_{k,n}$. Thus, to calculate the rate outage probability, it is required to find the probability density function (pdf) of $Y_k$ for a given allocated subcarriers' subset $\mathcal{C}_k$. To derive the pdf of $Y_k$, we first derive the moment generating function (MGF) $\Delta_{Y_k}(s)$ of the random variable $Y_k$ as illustrated by the following theorem:

**Theorem 1:** The MGF of $Y_k$ defined as a product of i.n.i.d Gamma distributed random variables is given by

$$\Delta_{Y_k}(s) = \frac{1}{\prod_{n \in \mathcal{C}_k} \Gamma(m_{k,n})} \quad (16)$$

$$\times G_{M_k,1}^{1,M_k}\left[-s \prod_{n \in \mathcal{C}_k} \mathsf{s}_{k,n} \,\Big|\, \begin{matrix} 1-m_{k,1},\ldots,1-m_{k,M_k} \\ 0 \end{matrix} \right]$$

where $G_{M_k,1}^{1,M_k}[.]$ is a Meijer G-function which is included as a standard built-in function in most of the mathematical software packages such as MATHEMATICA and MAPLE.

**Proof:** The MGF $\Delta_{Y_k}(s)$ is the expectation of $e^{sY_k}$, i.e.

$$\Delta_{Y_k}(s) = E\{e^{sY_k}\} \quad (17)$$

$$= \int_0^\infty \int_0^\infty \ldots \int_0^\infty e^{sx_1 x_2 \ldots x_{M_k}}$$

$$\times f_{X_{k,1}}(x_1) f_{X_{k,2}}(x_2) \ldots f_{X_{k,M_k}}(x_{M_k}) dx_1 \ldots dx_{M_k}$$

Define $J_i \triangleq x_i x_{i+1} \ldots x_{M_k}$ and

$$\prime_1 \triangleq \int_0^\infty x_1^{m_{k,1}-1} e^{-x_1/\mathsf{s}_{k,1}} e^{sx_1 J_2} dx_1 \quad (18)$$

which is the innermost integral after dropping the constant $1/\Gamma(m_{k,1})\mathsf{s}_{k,1}^{m_{k,1}}$, for simplicity. The integral in (18) can be rewritten in terms of the Meijer-G function [12] as:

$$\prime_1 = \int_0^\infty x_1^{m_{k,1}-1} G_{0,1}^{1,0}[\frac{x_1}{\mathsf{s}_{k,1}} \,\Big|\, \begin{matrix} - \\ 0 \end{matrix}] G_{0,1}^{1,0}[-sx_1 J_2 \,\Big|\, \begin{matrix} - \\ 0 \end{matrix}] dx_1 \quad (19)$$

Using the result provided in [13] for calculating integrals of hypergeometric type functions, one can find that the integral in (19) can be solved as:

$$\prime_1 = \mathsf{s}_{k,1}^{m_{k,1}} G_{1,1}^{1,1}\left[-s\mathsf{s}_{k,1} J_2 \,\Big|\, \begin{matrix} 1-m_{k,1} \\ 0 \end{matrix}\right] \quad (20)$$

The integration on $x_2$ defined as $\prime_2$ is given as:

$$\prime_2 \triangleq \int_0^\infty x_2^{m_{k,2}-1} e^{-x_2/\mathsf{s}_{k,2}} \prime_1 dx_2 \quad (21)$$

$$= \mathsf{s}_{k,1} \int_0^\infty x_2^{m_{k,2}-1} e^{-x_2/\mathsf{s}_{k,2}} \quad (22)$$

$$\times G_{1,1}^{1,1}\left[-s\mathsf{s}_{k,1} J_2 \,\Big|\, \begin{matrix} 1-m_{k,1} \\ 0 \end{matrix}\right] dx_2$$

$$= \mathsf{s}_{k,1} \int_0^\infty x_2^{m_{k,2}-1} e^{-x_2/\mathsf{s}_{k,2}} \quad (23)$$

$$\times G_{1,1}^{1,1}\left[-s\mathsf{s}_{k,1} J_3 x_2 \,\Big|\, \begin{matrix} 1-m_{k,1} \\ 0 \end{matrix}\right] dx_2$$

$$= \int_0^\infty x_2^{m_{k,2}-1} G_{0,1}^{1,0}[\frac{x_2}{\mathsf{s}_{k,2}} \,\Big|\, \begin{matrix} - \\ 0 \end{matrix}] \quad (24)$$

$$\times G_{1,1}^{1,1}\left[-s\mathsf{s}_{k,1} J_3 x_2 \,\Big|\, \begin{matrix} 1-m_{k,1} \\ 0 \end{matrix}\right] dx_2 \quad (25)$$

Similarly, using the result from [13], the integral in (21) is given by:

$$\prime_2 = \mathsf{s}_{k,1}^{m_{k,1}} \mathsf{s}_{k,2}^{m_{k,2}} G_{1,2}^{2,1}\left[-s\mathsf{s}_{k,1}\mathsf{s}_{k,2} J_3 \,\Big|\, \begin{matrix} 1-m_{k,1}, 1-m_{k,2} \\ 0 \end{matrix}\right] \quad (26)$$

Following the same argument used for calculating the second integral, the M-fold integral in (17), $\prime_{M_k}$, is given by induction as:

$$\prime_{M_k} = \left(\prod_{n \in \mathcal{C}_k} \mathsf{s}_{k,n}^{m_{k,n}}\right) \quad (27)$$

$$\times G_{1,M_k}^{M_k,1}\left[-s\prod_{n \in \mathcal{C}_k} \mathsf{s}_{k,n} \,\Big|\, \begin{matrix} 1-m_{k,1},\ldots,1-m_{k,M_k} \\ 0 \end{matrix}\right]$$

Given $\prime_{M_k}$, the closed-form formula of the MGF $\Delta_{Y_k}(s)$

is:

$$\Delta_{Y_k}(s) = \frac{s^{M_k}}{\prod_{n \in \mathcal{C}_k} \mathsf{s}_{k,n}^{m_{k,n}} \prod_{n \in \mathcal{C}_k} \Gamma(m_{k,n})} \quad (28)$$

Substituting (27) in (28) we get the result in (16).
The probability density function (pdf) of the random variable $Y_k$ is given in the following corollary.

**Corollary 1:** The pdf of $Y_k$ is given by:

$$f_{Y_k | \mathcal{C}_k}(y) = \frac{(y-1)^{-1} G_{0,M_k}^{M_k,0}\left[\frac{(y-1)}{\prod_{n \in \mathcal{C}_k} \mathsf{s}_{k,n}} \Big|_{m_{k,1}, m_{k,2}, \ldots, m_{k,M_k}}^{-}\right]}{\prod_{n \in \mathcal{C}_k} \Gamma(m_{k,n})}. \quad (29)$$

**Proof:** The pdf of $Y_k$ can be found as the inverse Laplace transform of $\Delta_{Y_k}(-s)$, that is

$$f_{Y_k | \mathcal{C}_k}(y) = \mathcal{L}^{-1}\left\{\Delta_{Y_k}(-s), y\right\} \quad (30)$$

Using the formula for the inverse Laplace transform of the Meijer G-function from [14], we obtain the result in (29).

**Corollary 2:** The cumulative distribution function (cdf), $F_{Y_k | \mathcal{C}_k}(y)$, of $Y_k$, is given by:

$$F_{Y_k | \mathcal{C}_k}(y) = \frac{G_{1,M_k+1}^{M_k,1}\left[\frac{(y-1)}{\prod_{n \in \mathcal{C}_k} \mathsf{s}_{k,n}} \Big|_{m_{k,1}, m_{k,2}, \ldots, m_{k,M_k}, 0}^{1}\right]}{\prod_{n \in \mathcal{C}_k} \Gamma(m_{k,n})} \quad (31)$$

**Proof:** By definition, the cdf $F_{Y_k | \mathcal{C}_k}(y)$ is the probability that the random variable $Y_k$ is in the interval $[1, y]$, that is

$$F_{Y_k | \mathcal{C}_k}(y) = \int_1^y f_{Y_k | \mathcal{C}_k}(z) dz \quad (32)$$

Therefore,

$$F_{Y_k | \mathcal{C}_k}(y) = \int_0^{y-1} \frac{G_{0,M_k}^{M_k,0}\left[\frac{z}{\prod_{n \in \mathcal{C}_k} \mathsf{s}_{k,n}} \Big|_{m_{k,1}, m_{k,2}, \ldots, m_{k,M_k}}^{-}\right]}{z \prod_{n \in \mathcal{C}_k} \Gamma(m_{k,n})} dz$$

$$= \frac{1}{\prod_{n \in \mathcal{C}_k} \Gamma(m_{k,n})} \frac{1}{2\pi j} \int_{\mathcal{U}} \prod_{n \in \mathcal{C}_k} \Gamma(m_{k,n} + \ddagger)$$

$$\times \left(\frac{1}{\prod_{n \in \mathcal{C}_k} \mathsf{s}_{k,n}}\right)^{-\ddagger} \int_0^{y-1} z^{-\ddagger - 1} dz d\ddagger$$

$$= \frac{1}{\prod_{n \in \mathcal{C}_k} \Gamma(m_{k,n})} \frac{1}{2\pi j} \int_{\mathcal{U}} \prod_{n \in \mathcal{C}_k} \Gamma(m_{k,n} + \ddagger)$$

$$\times \left(\frac{1}{\prod_{n \in \mathcal{C}_k} \mathsf{s}_{k,n}}\right)^{-\ddagger} \frac{(y-1)^{-\ddagger}}{-\ddagger} d\ddagger$$

$$= \frac{1}{\prod_{n \in \mathcal{C}_k} \Gamma(m_{k,n})} \frac{1}{2\pi j} \int_{\mathcal{U}} \prod_{n \in \mathcal{C}_k} \Gamma(m_{k,n} + \ddagger)$$

$$\times \frac{\Gamma(-\ddagger)}{\Gamma(1-\ddagger)} \left(\frac{(y-1)}{\prod_{n \in \mathcal{C}_k} \mathsf{s}_{k,n}}\right)^{-\ddagger} d\ddagger \quad (33)$$

where in the last step in (33) we used the equality $\Gamma(1-\ddagger) = -\ddagger \Gamma(-\ddagger)$.
$\mathcal{U}$ is an appropriate integration contour, and $j = \sqrt{-1}$. Solving the integral in (33) will yield the solution in (31).

Using the CDF, $F_{Y_k}(y) = Pr(Y_k \leq y)$, in (31) and (4), the closed form of the rate outage probability for a given $\mathcal{C}_k$ is given by:

$$P_k^{out} = F_{Y_k | \mathcal{C}_k}(2^{r_{k,min}/B_{sc}}) \quad (34)$$

$$= \frac{G_{1,M_k+1}^{M_k,1}\left[\frac{(2^{r_{k,min}/B_{sc}}-1)}{\prod_{n \in \mathcal{C}_k} \mathsf{s}_{k,n}} \Big|_{m_{k,1}, m_{k,2}, \ldots, m_{k,M_k}, 0}^{1}\right]}{\prod_{n \in \mathcal{C}_k} \Gamma(m_{k,n})}.$$

As an indicative example for the calculated outage probability, the outage probability is plotted in Figure 3 as a function of the minimum required data rate per user for different values of the channel physical bandwidth. The observed results evidently show that the outage probability decreases with the increase of the channel physical bandwidth for the same required data rate. Moreover, For larger channel bandwidths, for example when B is 10 MHz and 5 MHz, the difference in the outage probability values become tighter, as shown in Figure 3, with outage probability values of $2 \times 10^{-4}$ and $2 \times 10^{-5}$ for a bandwidth of 5MHz and 10MHz respectively.

We performed simulation experiments based on the system model described in Section 2 to compare the calculated rate outage probability in (34) against the rate outage probability evaluated from the simulation. The network consists of one BS and multiple SS. The BS allocates the available SCs per user based on the users' data rate requirement and their channel quality. This process is repeated each frame duration. For our simulation, we used a frame size of 10 ms. The matrix of SCs is generated every frame based on the Nakagami-m channel model and the SCs are allocated per user to meet the users minimum data rate requirements. The attainable data rate is calculated given the allocated SCs per user. The simulated outage probability is calculated as the difference between the minimum required data rate and the attainable data rate. The calculated outage probability is obtained from (34) by substituting the bandwidth of the same SCs used in simulation per user and the Nakagami-m pdf parameters. Results are shown in Figure 4 for the calculated outage probability against the simulated outage probability for different values of the minimum required data rates. The results in Figure 4 clearly show that the rate calculated outage probability matches that obtained from the simulation for different values of the user required data rate. Hence, the calculated rate outage probability is verified to provide a closed form expression of the rate outage probability in an OFDMA system.

## IV. DERIVATION OF MULTI-HOP RATE OUTAGE PROBABILITY

In this Section, using the results of section 3, we derive the rate outage probability of OFDMA multi-hop communication system under Nakagami-m channel models.

For multi-hop scenario, a minimum requested transmission rate requirement for a given SS is satisfied when the minimum supported transmission rate on an intermediate link (bottle-neck) in the routing path between the BS and this particular SS is greater than the requested rate. Otherwise, a rate outage will occur. Therefore, the multi-hop rate outage probability is the probability that the supported transmission rate on the bottle-neck link is less than the minimum requested transmission rate by a particular SS. In order to mathematically express this probability, let us define the followings. For the intermediate link (hop) $i$, $i \in \{1, 2, \cdots, N_{h,k}\}$ with $N_{h,k}$ is the number of hops (distance) between the BS and kth SS, let

$$Y_k^i \triangleq \prod_{n \in \mathcal{C}_k^i} X_{k,n}, \quad (35)$$

where $\mathcal{C}_k^i$ is the subset of data subcarriers allocated to carry the data of the kth SS on the ith hop. Henceforth, the achievable rate per OFDMA symbol of the $k$ th SS $r_k^i$ on $i$th hop is:

$$r_k^i = \frac{B}{N} \log_2(Y_k^i) \quad (36)$$

The minimum supported transmission rate on the route from the BS to the SS $k$ is defined as

$$r_k^{bn} = min\{r_k^1, r_k^2, \cdots, r_k^{N_{h,k}}\} \quad (37)$$

From the above, we can figure out that the multi-hop rate outage probability of the kth SS, $P_k^{out-mh}$, is given by

$$P_k^{out-mh} = P_r\left(r_k^{bn} \mid \mathcal{C}_k < r_{k,min}\right) \quad (38)$$

or

$$P_k^{out-mh} = P_r\left(Y_k^{bn} \mid \mathcal{C}_k < 2^{r_{k,min}/B_{sc}}\right) \quad (39)$$

Here,

$$\mathcal{C}_k = \bigcup_{i=1}^{N_{h,k}} \mathcal{C}_k^i$$

Assuming that $\{Y_k^i\}_{i=1}^{N_{h,k}}$ are independent random variables, then

$$P_k^{out-mh} = 1 - P_r\left(Y_k^{bn} \mid \mathcal{C}_k \geq 2^{r_{k,min}/B_{sc}}\right), \quad (40)$$

$$= 1 - \prod_{i=1}^{N_{h,k}} P_r\left(Y_k^i \mid \mathcal{C}_k^i \geq 2^{r_{k,min}/B_{sc}}\right)$$

$$= 1 - \prod_{i=1}^{N_{h,k}} \left(1 - P_r\left(Y_k^i \mid \mathcal{C}_k^i < 2^{r_{k,min}/B_{sc}}\right)\right)$$

Note that $P_r\left(Y_k^i \mid \mathcal{C}_k^i < 2^{r_{k,min}/B_{sc}}\right)$ is the rate outage probability at the ith hop, i.e. $P_k^{out,i}$, thus (40) can be written as

$$P_k^{out-mh} = 1 - \prod_{i=1}^{N_{h,k}} \left(1 - P_k^{out,i}\right), \quad (41)$$

with

$$P_k^{out,i} = \frac{G_{1,M_{k,i}+1}^{M_{k,i},1}\left[\frac{(2^{r_{k,min}/B_{sc}}-1)}{\prod_{n \in \mathcal{C}_k^i} s_{k,n}} \middle| \begin{array}{c} 1 \\ m_{k,1}, m_{k,2}, \cdots, m_{k,M_{k,i}}, 0 \end{array} \right]}{\prod_{n \in \mathcal{C}_k^i} \Gamma(m_{k,n})}$$

where $M_{k,i} = |\mathcal{C}_k^i|$, i.e. the number of data subcarriers allocated to carry the data of SS $k$ on the ith hop. It should be noted that the assumption of the independence of $\{Y_k^i\}_{i=1}^{N_{h,k}}$ can result from assigning each subcarrier once in the whole network, or from the multiuser diversity even if the same subcarrier has been assigned again in the network after some hops (to eliminate interference).

In line with the single-hop simulation results provided in 4, we performed simulation experiment to compare the calculated multi-hop rate outage probability in (41). We used the same environment and simulation parameters as that used for the single hop. Our simulation is based on Type-II LTE-relay and the transparent 802.16j relay mode. We chose these modes to allow centralized SCs allocation at the BS. Type-I LTE-relay and non-transparent 802.16j relay mode do not entail much difference from the point of view of verifying the correctness of the outage probability derivation. Note that we derived the closed form rate outage probability for a given user assuming the allocated sub-carriers' set is pre - allocated. The algorithm for allocating subcarriers to network users is out of the scope of this work. Algorithms for allocating subcarriers could be performed by the relay node in the non-transparent mode (distributed SCs allocation) or by the BS (centralized SCs allocation). For the sake of simplicity, we simulate here the transparent relay mode. Transparent relay mode restricts the number of hops in the network into 2 hops. The BS allocates SCs for the relay node and the SS given the Nakagami-m channel condition at each subcarrier at the relay and the access link. The attainable rate of the allocated SCs at each link is compared to the minimum required data rate to identify the bottleneck link. The simulated outage probability is calculated as the difference between the required data rate and the bottleneck attainable data rate. The calculated outage probability is calculated from (41) given the allocated SCs used in simulation and the Nakagami-m pdf parameters. Results are shown in Figure 5. Figure 5 shows similar results to those obtained for single-hop outage probability. The calculated results almost match the simulated results for low and high required data rates. Hence, we can safely deduce that equation (41) provides a closed form expression of the rate outage probability in multi-hop OFDMA systems.

V. OUTAGE PROBABILITY APPLICATIONS

The closed form of the rate outage probability can be utilized in several aspects. For example, it can be integrated into a subcarrier allocation scheme, where the BS allocates the subcarriers and subcarriers' power level to the active SSs in the system to achieve a requested outage probability bound. Also, it can be used as a performance measure to evaluate the efficiency of different resource allocation algorithms, or to improve the performance of a certain resource allocation algorithm through adapting the resource allocation based on the feedback information of the channel status and the outage probability of users. As a final example, The SS rate outage probability can be used to calculate the network rate outage probability which can be used latter in efficient dynamic admission control algorithms.

In this Section, as an example, we consider a centralized subcarrier allocation scheme that is aware of the outage probability of each user, called **SCA-OUT**, for single hop cellular network. In **SCA-OUT**, the BS allocates the subcarriers to the $K$ active SSs in the system and then loads each SC with an optimal power level such that the overall DL achievable transmission rate is maximized. The constraint can be the exponential average transmission rate. This is assured to be bounded from above and from below by a given probability. Mathematically speaking, the problem is formulated as follows:

$$\text{SCA-OUT: } \max_{p_{k,n}, \mathcal{C}_k} \left\{ B_{sc} \sum_{k=1}^{K} \sum_{n \in C_k} \log_2(1+\mathsf{x}_{k,n}(t)) \right\} \quad (42)$$

s.t.:

$$Pr\left(r_{k,min} \leq R_k(t) \leq \varpi_k r_{k,min} \mid \mathcal{C}_k\right) \geq \mathsf{v} \quad (43)$$
$$p_{k,n} \geq 0, \mathcal{C}_k \neq \varnothing$$

where $1 \leq \varpi_k \leq r_{k,max}/r_{k,min}$ and

$$R_k(t) = (1-\frac{1}{T})R_k(t-1) + \frac{1}{T}r_k(t), \quad (44)$$

is the exponential average received rate over a window size of $T$ which is an integer multiple of the time frame $T_f$. Using the fact that $R_k(-1) = 0$ one can rewrite (44) as follows:

$$R_k(t) = \frac{1}{T}\sum_{n=0}^{t}(1-\frac{1}{T})^{t-n} r_k(n) \quad (45)$$

From (9), the achievable rate at the $j$ th frame of the $k$ th SS, $r_k(j)$ can be expressed as follows:

$$r_k(j) = \frac{B}{N}\log_2(Y_k(j)) \quad (46)$$
$$= \frac{B}{N}\log_2(1+\mathsf{x}_{k,eff}(j)) \quad (47)$$

where $\mathsf{x}_{k,eff}(j) = Y_k(j) - 1$ is the effective SNR over all $k$ th SS's allocated subcarriers. Henceforth, $\mathsf{x}_{k,eff}(j)$ is a random variable with the following probability density function:

$$f_{\mathsf{x}_{k,\text{eff}}(j)|\mathcal{C}_k}(y) = \frac{(y)^{-1} G_{0,M_k}^{M_k,0}\left[\frac{y}{\prod_{n\in\mathcal{C}_k}\overline{\mathsf{x}}_{k,n}}\Bigg|_{m_{k,1},m_{k,2},\ldots,m_{k,M_k}}^{-}\right]}{\prod_{n\in\mathcal{C}_k}\Gamma(m_{k,n})}$$

$$\forall y \geq 0, \qquad (48)$$

and accordingly, the CDF is given by:

$$F_{\mathsf{x}_{k,\text{eff}}(j)|\mathcal{C}_k}(y) = \frac{G_{1,M_k+1}^{M_k,1}\left[\frac{y}{\prod_{n\in\mathcal{C}_k}\overline{\mathsf{x}}_{k,n}}\Bigg|_{m_{k,1},m_{k,2},\ldots,m_{k,M_k},0}^{1}\right]}{\prod_{n\in\mathcal{C}_k}\Gamma(m_{k,n})}, \qquad (49)$$

To be able to solve the optimization problem in **SCA-OUT**, we need to provide a closed form expression of the probability given in (42). Therefore, we need to find the pdf of $R_k(t)$, which is a weighted sum of independent random variables $\{r_k(n)\}_{n=0}^{n=t}$. Note that the independence of the random variables $\{r_k(n)\}_{n=0}^{n=t}$ is a result of the assumption that the channel gains over all subcarriers between the $k$ th SS and BS only change on a time frame basis. Next, we introduce some preliminary results that lead us to an approximate closed form formula of the constraint in (43).

With the deployment of the AMC module at both the BS and SS to keep the bit error rate (BER) at a prescribed value, the achievable rate is a discrete random variable, see Table 1. Henceforth, the probability mass function (pmf) of $r_k(n)$ at the $n$ th time frame denoted to $\Psi_i$, and is given by:

$$\Psi_i \triangleq P_r(r_k(n) = \epsilon_i) \qquad (50)$$
$$= P_r(SNR_i \leq \mathsf{x}_{k,\text{eff}}(n) < SNR_{i+1}) \qquad (51)$$
$$= \int_{SNR_i}^{SNR_{i+1}} f_{\mathsf{x}_{k,\text{eff}}(j)|\mathcal{C}_k}(z)dz$$
$$= F_{\mathsf{x}_{k,\text{eff}}(j)|\mathcal{C}_k}(SNR_{i+1}) - F_{\mathsf{x}_{k,\text{eff}}(j)|\mathcal{C}_k}(SNR_i)$$

where $\{SNR_i\}_{i=1}^{7}$ are the thresholds of the 7 non-overlapping regions shown in Table 1 with $\mathsf{x}_8 = \infty$, $\mathsf{x}_0 = 0$, and $\epsilon_i \in \{0.5, 1, 1.5, 2, 3, 4, 4.5\}$ is the number of information bits on the $M$-ary QAM symbol. Providing a closed formula for the probability in (42) requires the probability density function (pdf) of $R_k(t)$. Assuming that first and second order moments of $R_k(t)$ are finite, for a large $t$, and according to the Central Limit Theorem (CLT), the pdf of $R_k(t)$ can be approximated as a Normal distributed random variable as follows:

$$\frac{R_k(t) - \tilde{}_{R_k(t)}}{\dagger_{R_k(t)}} \to \mathcal{N}(0,1), \qquad (52)$$

where $\mathcal{N}(0,1)$ denotes a normal distribution with zero mean and variance equal to 1, with $\tilde{}_{R_k(t)}$ being the mean of $R_k(t)$ and $\dagger^2_{R_k(t)}$ is the variance of $R_k(t)$. Both $\tilde{}_{R_k(t)}$ and $\dagger^2_{Rk(t)}$ can be found as functions of $\tilde{}_{r_k(n)}$ and $\dagger^2_{r_k(n)}$ which are the mean and the variance of $r_k(n)$, respectively:

$$\tilde{}_{Rk(t)} = \frac{1}{T}\sum_{n=0}^{t}(1-\frac{1}{T})^{t-n}\tilde{}_{r_k(n)} \qquad (53)$$
$$= \frac{1}{T}\sum_{n=0}^{t}(1-\frac{1}{T})^{t-n}\sum_{i=1}^{7}\epsilon_i \Psi_i$$

and

$$\dagger^2_{R_k(t)} = \left(\frac{1}{T}\right)^2 \sum_{n=0}^{t}(1-\frac{1}{T})^{2(t-n)}\dagger^2_{r_k(n)}$$

where,

$$\dagger^2_{r_k(n)} = E\{r_k^2(n)\} - \tilde{}^2_{r_k(n)} \qquad (54)$$
$$= \sum_{i=1}^{7}\epsilon_i^2 \Psi_i - \left(\sum_{i=1}^{7}\epsilon_i \Psi_i\right)^2$$

Therefore, the approximate pdf of the random variable $R_k(t)$ is given as:

$$f_{R_k(t)}(z) = \frac{1}{\sqrt{2f}\,\dagger_{R_k(t)}} exp\left(-\frac{(z - \tilde{}_{R_k(t)})^2}{2\dagger^2_{R_k(t)}}\right) \qquad (55)$$

Next we rewrite an approximate closed form expression for equation (42) as follows:

$$Pr(r_{k,\min} \leq R_k(t) \leq \ldots_k r_{k,\min} | \mathcal{C}_k)$$
$$= \int_{r_{k,\min}}^{\ldots_k r_{k,\min}} f_{R_k(t)}(z)dz \qquad (56)$$

$$= \frac{1}{2}\left(erf\left(\frac{\ldots r_{k,\min} - \tilde{}_{R_k(t)}}{\sqrt{2}\,\dagger_{R_k(t)}}\right) - erf\left(\frac{r_{k,\min} - \tilde{}_{R_k(t)}}{\sqrt{2}\,\dagger_{R_k(t)}}\right)\right)$$

where $erf(\Theta) = \frac{2}{\sqrt{f}}\int_0^{\Theta}\exp(-x^2)dx$ is the error function. Then, the problem **SCA-OUT** can be rewritten as follows:

$$\textbf{SCA-OUT:} \max_{p_{k,n}, \mathcal{C}_k} \left\{ B_{sc}\sum_{k=1}^{K}\sum_{n\in\mathcal{C}_k}\log_2(1+\mathsf{x}_{k,n}(t))\right\} \qquad (57)$$

$$s.t.: \quad \frac{1}{2}\left(erf\left(\frac{r_{k,min} - \tilde{R}_k(t)}{\sqrt{2}\dagger_{R_k(t)}}\right)\right.$$

$$\left. - erf\left(\frac{r_{k,min} - \tilde{R}_k(t)}{\sqrt{2}\dagger_{R_k(t)}}\right)\right) \geq \vee$$

$$p_{k,n} \geq 0, \mathcal{C}_k \neq \emptyset \quad (58)$$

This problem can be either solved analytically or by a heuristic algorithm, which is beyond the scope of the paper.

## VI. CONCLUSIONS

We introduced an analytical model to calculate a tractable closed form rate outage probability expression of a SS in a single and multi-hop cellular network. The closed form expression of the rate outage probability is obtained by analyzing the desired data rate of each user against the link attainable transmission rate. The problem is formulated as a function of SNR per OFDMA subcarrier and the Nakagami-m channel model. Simulation results verified the match of the calculated outage probability with the simulated one in an OFDMA system. Moreover, we formulated an optimization problem for single-hop scenarios in which subcarriers are assigned to admitted users in the network such that the total transmission rate is maximized while catering for fairness for all users. Fairness was achieved by guaranteeing a minimum average rate outage probability over a time window of length equal to integer multiple of the time frame for each admitted user.

ACKNOWLEDGMENT

This work was made possible by NPRP grant # NPRP4-553-2-210 from the Qatar National Research Fund (a member of The Qatar Foundation). The statements made herein are solely the responsibility of the authors

Table 1: Modulation and Coding Schemes for IEEE 802.16.

| Modulation (coding) | Info bits/symbol | Required SNR |
|---|---|---|
| BPSK(1/2) | 0.5 | 6.4 |
| QPSK(1/2) | 1 | 9.4 |
| QPSK(3/4) | 1.5 | 11.2 |
| QAM(1/2) | 2 | 16.4 |
| QAM(3/4) | 3 | 18.2 |
| QAM(2/3) | 4 | 22.7 |
| QAM(3/4) | 4.5 | 24.4 |

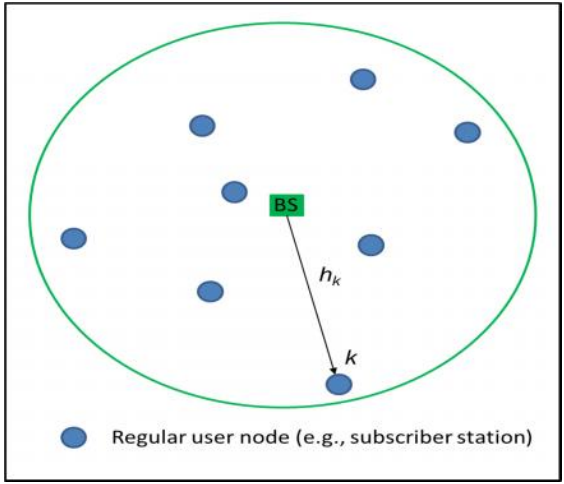

Figure 1: Single hop cellular radio network

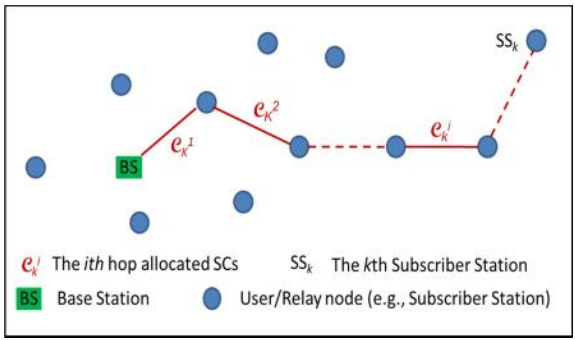

Figure 2: Multi-hop cellular radio network abstraction

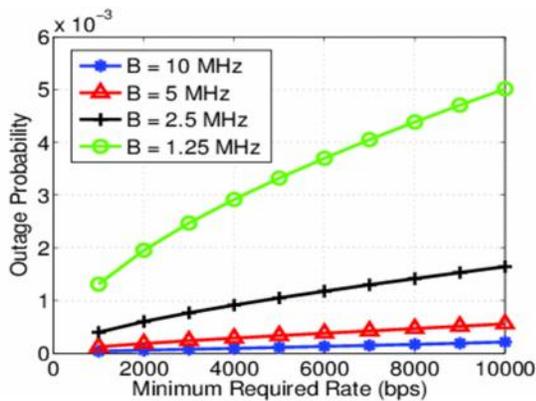

Figure 3: Single hop Outage Probability against the Required Data Rate for different channel bandwidth

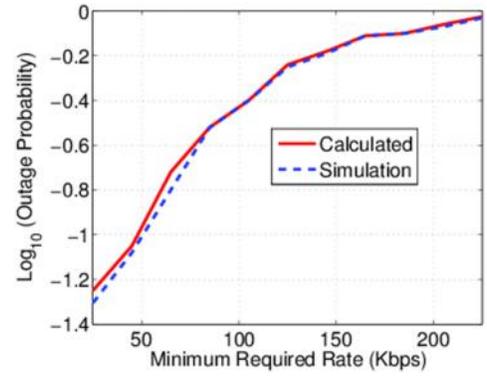

Figure 4: Single hop Outage Probability against the Required Data Rate, calculated and simulated

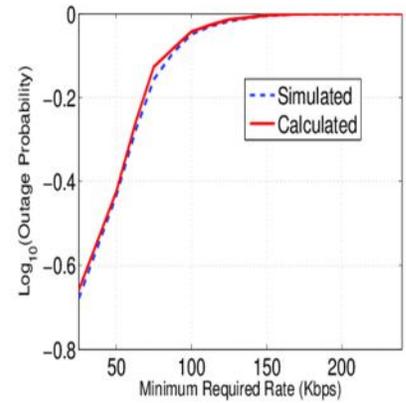

Figure 5: Multi-hop Outage Probability against the Required Data Rate, calculated and simulated